\documentclass[letterpaper]{article} % DO NOT CHANGE THIS
\usepackage[submission]{aaai25}  % DO NOT CHANGE THIS
\usepackage{times}  % DO NOT CHANGE THIS
\usepackage{helvet}  % DO NOT CHANGE THIS
\usepackage{courier}  % DO NOT CHANGE THIS
\usepackage[hyphens]{url}  % DO NOT CHANGE THIS
\usepackage{graphicx} % DO NOT CHANGE THIS
\usepackage{natbib}  % DO NOT CHANGE THIS AND DO NOT ADD ANY OPTIONS TO IT
\usepackage{caption} % DO NOT CHANGE THIS AND DO NOT ADD ANY OPTIONS TO IT
\usepackage{algorithm}
\usepackage{algorithmic}
\usepackage{booktabs} 
\usepackage{bm}
\usepackage{multirow}
\usepackage{makecell}
\newcommand{\frameworkname}{DFT} 
\usepackage{tabularx}
\usepackage{subcaption}
\usepackage{amsthm}
\usepackage{amssymb}
\usepackage{amsmath}
\usepackage{newfloat}
\usepackage{listings}
\DeclareCaptionStyle{ruled}{labelfont=normalfont,labelsep=colon,strut=off} % DO NOT CHANGE THIS
\floatstyle{ruled}
\newfloat{listing}{tb}{lst}{}
\floatname{listing}{Listing}
\title{\frameworkname: A Dual-branch Framework of Fluctuation and Trend for Stock Price Prediction}
\author{
   Chengqi Dong, Zhiyuan Cao, S Kevin Zhou, Jia Liu
}
\affiliations{

}

\usepackage{bibentry}
\begin{document}

\maketitle

\begin{abstract}
Stock price prediction is of significant importance in quantitative investment. Existing approaches encounter two primary issues: First, they often overlook the crucial role of capturing short-term stock fluctuations for predicting high-volatility returns. Second, mainstream methods, relying on graphs or attention mechanisms, inadequately explore the temporal relationships among stocks, often blurring distinctions in their characteristics over time and the causal relationships before and after. However, the high volatility of stocks and the intricate market correlations are crucial to accurately predicting stock prices. To address these challenges, we propose a Dual-branch Framework of Fluctuation and Trend (\frameworkname), which decomposes stocks into trend and fluctuation components. By employing a carefully design decomposition module, \frameworkname~effectively extracts short-term fluctuations and trend information from stocks while explicitly modeling temporal variations and causal correlations. Our extensive experiments demonstrate that \frameworkname~outperforms existing methods across multiple metrics, including a 300\% improvement in ranking metrics and a 400\% improvement in portfolio-based indicators. Through detailed experiments, we provide valuable insights into different roles of trends and fluctuations in stock price prediction. Code is available at \url{https://github.com/cq-dong/DFT_25}. 
\end{abstract}


\section{Introduction}

Stock price prediction is a fundamental task in the field of quantitative investment~\cite{stockmixer}. However, predicting stock price trends is extremely challenging due to the high volatility and chaos of the stock market. Active trading behaviors such as buying and selling by investors drive stock price fluctuations. In addition, the stock market is also affected by many factors, including economic indicators, financial reports, political events, investor sentiment, etc.~\cite{mdgnn}. Achieving a high prediction accuracy remains an ongoing challenge in this domain.

\begin{figure}[htbp]
\centering
     \includegraphics[width=1.0\linewidth]{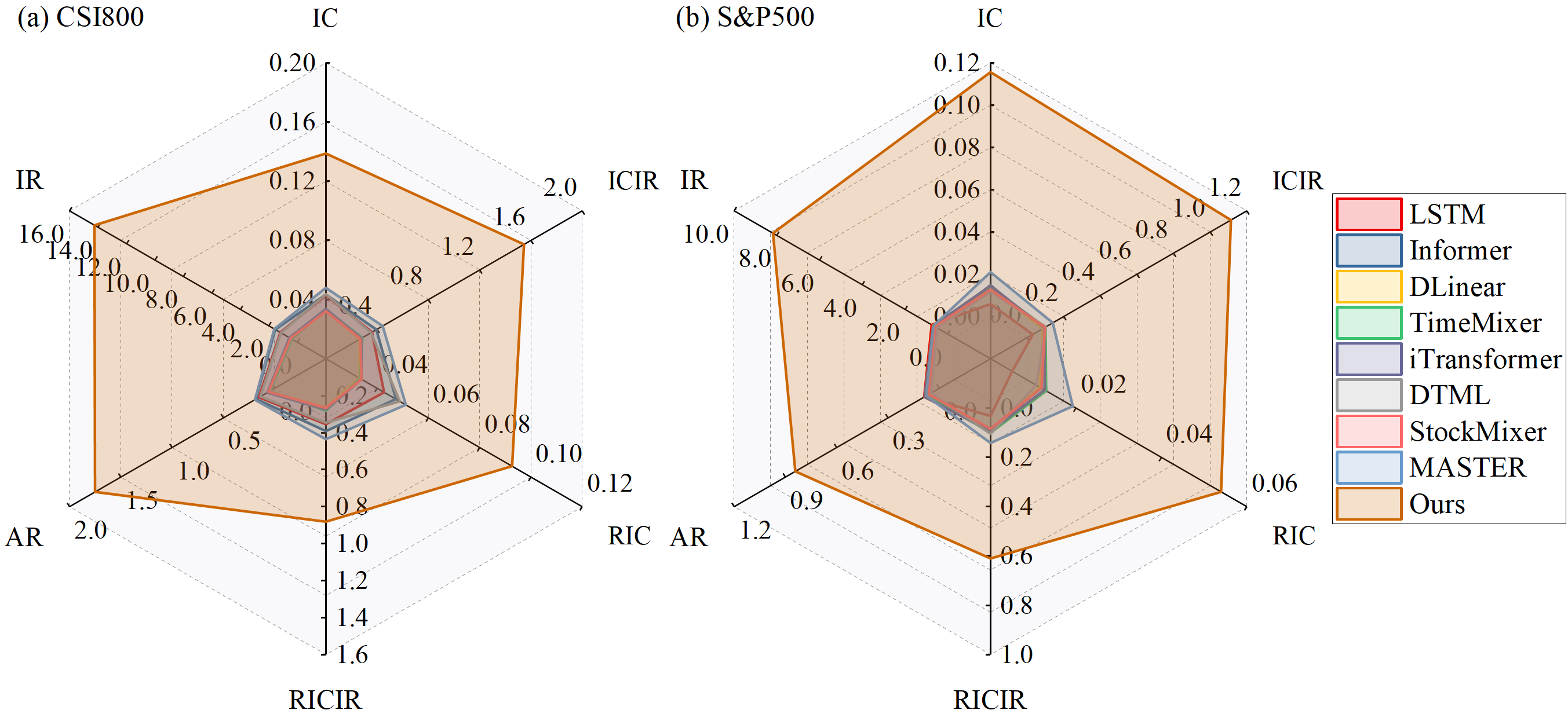}
\caption{Overall performance comparsion on the CSI800 and S\&P500 stock datasets.}
\label{fig:spider1}
% \vspace{-10px}
\end{figure}

Many works have achieved remarkable results in improving prediction performance. Traditional machine learning methods, such as decision trees and support vector machines, are used to model stock return changes~\cite{nugroho2014decision,chen2016xgboost,kamble2017short,xie2013semantic}. However, these methods require the manual construction of large financial indicator features, and it is difficult to model complex dynamic correlations between stocks.

With the advancement of deep learning, existing methods use the powerful representation ability of neural networks to mine stock trends and correlations. At present, there are two main research ideas for stock return prediction. (1) Time correlation. Stock price movements are caused by continuous changes in supply and demand, and time trend changes have obvious dependencies. Examples include recurrent neural networks~\cite{lstm2017stock,gru2014learning}, convolutional neural networks~\cite{conv2018empirical}, which model stock trends through individual stock time series features. (2) Stock correlation. Different stocks in the market form complex and dynamic dependencies due to various factors such as division of labor and industry status, and stock price fluctuations will affect each other. Fusion of stock spatiotemporal features through graphs or self-attention mechanisms~\cite{mdgnn,ci-sthpan, master} can effectively improve prediction capabilities. However, these methods face two limitations.

First, existing works ignore the importance of mining individual stock fluctuation information to predict stock returns. Stock trend information is a long-term, relatively stable change, representing the overall direction; stock fluctuation information refers to the short-term change component after removing the long-term trend, representing the profitability of the stock itself and reflecting the change in the power comparison of the long-short game. Fully exploiting volatility information can capture the implied price-volume patterns from the individual stock perspective and enhance forecasting ability when combined with the overall trend. Previous works often treat stock information as a whole without distinguishing between trend and fluctuation components, or they perform a simple decomposition followed by fusion and embedding, which leads to the mutual interference of trends and fluctuations between different stocks in the market.

Second, they often oversimplify the time correlation of stocks. Specifically, the correlation of stocks is dynamically updated rather than fixed, and the features of different time steps have time series feature differences. In addition, the interaction of stock representations depends on the time order. The information on earlier steps will affect the information on later steps and vice versa. However, when conducting correlation mining, existing methods either perform time alignment operations~\cite{dtml2021accurate}, which oversimplifies the unique properties of individual stocks at different time steps, or fail to distinguish the temporal representation of stocks, which ignores the causal relationship before and after the stock information~\cite{master}, weakening the model's time modeling ability.

To solve the above problems, we propose a \textbf{\underline{D}}ual-branch Framework of \textbf{\underline{F}}luctuation and \textbf{\underline{T}}rend (\frameworkname) for Stock Price Prediction. We believe that fluctuation information is a unique attribute of individual stocks, while trends are the concentrated expression of many stocks. Therefore, we design a fluctuation-trend dual-branch model to extract features separately. By separately modeling the order of correlation, we avoid the interference of fluctuation between stocks, and can give a full play to the different roles of the two in capturing stock characteristics, which is more conducive to prediction returns. In addition, we skillfully combine the advantages of RWKV time series representation and the relationship modeling ability of the self-attention mechanism while preserving the attributes of different time steps, maintaining the causal correlation of time series, and fully mining the complex cross-time causal relationship and stock correlation.
Our contributions are summarized as follows:
\begin{itemize}
\item We propose a new dual-branch stock prediction framework to effectively capture the temporal and stock correlations from two aspects: fluctuation and trend. To the best of our knowledge, this is the first stock price prediction model that uses a learning-based method to capture long-term trends and short-term fluctuation information simultaneously.
\item We consider the role of differentiated representations of different time steps in time series correlation and causal correlation and fully mine the complex time dependency in stock data.
\item We conducted experiments to validate the designs of our proposed method and demonstrated its superiority compared to baselines. Through a comprehensive analysis, we provide valuable insights into the different roles of trends and fluctuations in stock price forecasting and the impact of temporal causality.
\end{itemize}

\section{Related work}

\subsubsection{Stock Trend Prediction.}
In quantitative trading, the ability to predict stock trends is crucial. To achieve this, a multi-factor model is commonly employed~\cite{mdgnn}. This model considers multiple influencing factors from an econometric perspective, including trading volume, price dynamics, and fundamental company data such as earnings and debt ratios. With the popularity of deep learning, recurrent neural networks \cite{lstm2017stock, feng2019temporal} and convolutional neural networks \cite{conv2018empirical} have been utilized for short-term trend prediction. To enhance sensitivity to fine-grained transition signals, researchers have explored additional techniques such as adversarial training~\cite{stockformer} and gated causal convolution~\cite{stockmixer}.

\subsubsection{Stock Correlation Prediction.}
Modeling stock correlations is a key research focus in stock prediction. Stock price movements are influenced by both historical trends and related stock movements. Therefore, current research addresses stock connections from two perspectives: time series correlations and stock correlations. Researchers propose constructing static graphs of stock relationships based on industry sectors~\cite{ci-sthpan}, concepts~\cite{ijcai2020p626}, and more. Moreover, to adapt flexibly to market and stock changes, scholars employ hyper-graphs~\cite{liu2019transformer, ding2020hierarchical}, attention mechanisms~\cite{yoo2021accurate, xiang2022temporal}, and similar techniques for dynamic modeling of stock relationships.

\section{Methodology}

\subsection{Overview}
\begin{figure*}[htbp]
\centering
     \includegraphics[width=1.0\linewidth]{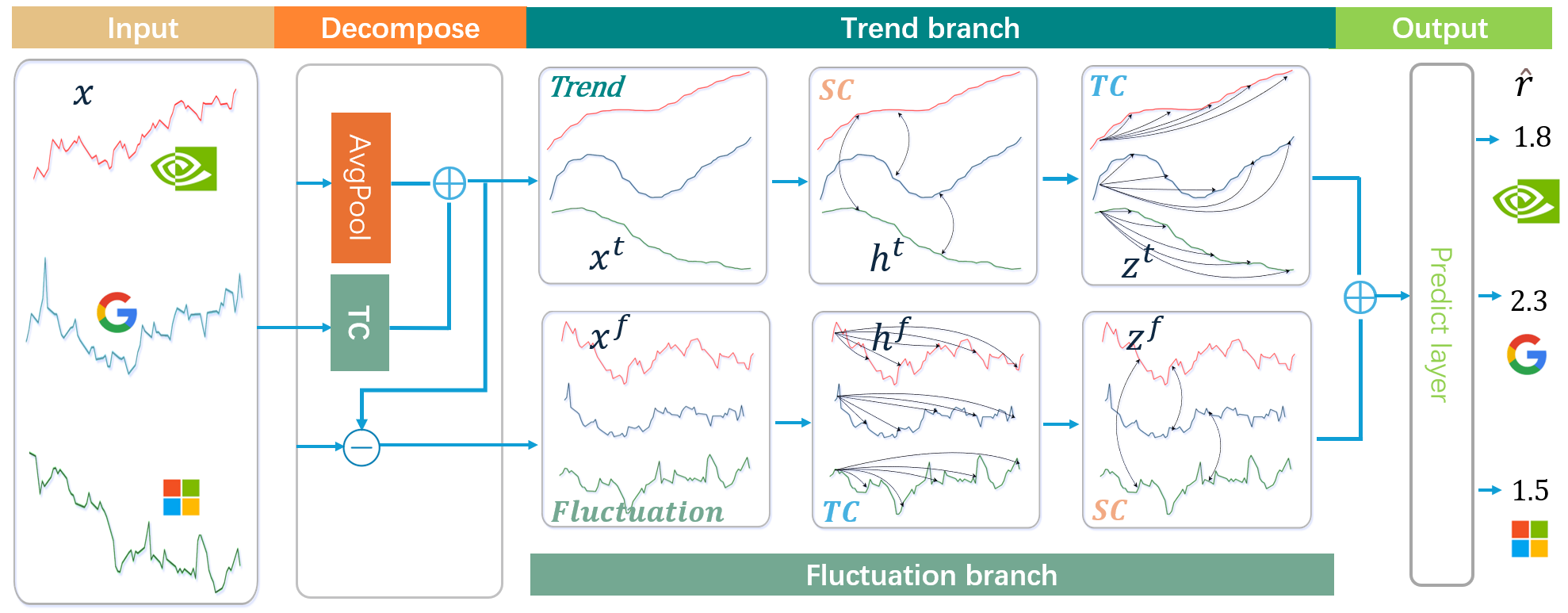}
\caption{Our model \frameworkname~ structure, a framework for independently modeling the correlation of time and stock dimensions in fluctuation and trend information.}
\label{fig:pdf}
\end{figure*}

In this section, we first give the definition of the stock price prediction task in this paper. Then, the proposed \frameworkname~ model is presented, as shown in Figure~\ref{fig:pdf}. The framework mainly includes three steps: (1) Stock decomposition. We design a decomposition module to obtain the overall trend information and daily fluctuation information of each stock in the viewing window $T$. (2) Correlation modeling. Taking into account the different characteristics of fluctuation and trend information mentioned above, we design an independent branch containing time and stock correlation for modeling. (3) Output prediction. We aggregate the time step information output by the two branches and input it into the prediction layer to obtain the prediction results.

\subsection{Problem definition}
Following existing works on stock market analysis \cite{sawhney2020spatiotemporal, huynh2023efficient, master}, for the stock price prediction problem, we focus on predicting the stock return ratio because it normalizes the market price variety between different stocks compared to the absolute price change. Specifically, for each time step of $t\in[1,T]$, we collect the features $\tilde{x}_{u,t}\in\mathbb{R}^{F}$ of each stock $u\in\mathcal{S}$, and the return ratio is expressed as $\tilde{r}_u=(c_{u,t+d}-c_{u,t+1})/c_{u,t+1}$, where $c_{u,t}$ represents the closing price of stock $u$ on day $t$, so the return ratio here represents the relative change of the closing price within $d$ days. Consistent with ~\cite{yang2020qlib}, we perform Z-score normalization on $\tilde{r}_u$ to obtain the label $r_{u}=\operatorname{Norm}_{\mathcal{S}}(\tilde{r}_{u})$ with ranking. According to the above definition, the stock prediction problem can be formalized as follows: Given stock features
$\{\tilde{x}_{u,t}\}_{u\in\mathcal{S},t\in[1,T]}$, stock price prediction is to jointly forecast the future normalized return ratio $\{r_u\}_{u\in\mathcal{S}\cdot}$

\subsection{Stock decomposition}
\subsubsection{Input embedding layer.} For each stock $\tilde{x}_{u,t} \in \mathbb{R}^{F}$, $u \in \mathcal{S},\; t \in [1,T]$, we employ a fully connected layer to obtain the embedding representation, which is $\mathcal{X}:\{{x}_{u,t}\}_{u\in\mathcal{S},t\in[1,T]}=FFN(\tilde{x}_{u,t})\in\mathbb{R}^{D}$.

\subsubsection{Information decomposition module.} Previous studies~\cite{stockformer,ci-sthpan} often ignore the fact that stock prices are affected by recent trends and daily fluctuations. We emphasize the importance of decomposing information into trends and fluctuations, especially the importance of fluctuations information for the final return prediction $r_{u}$. We first extract the trend component $\mathcal{X}^{t}$ from embedding $\mathcal{X}$. 
Considering that trend information is a long-term and stable feature, We decompose the input using AvgPool with kernel size $k_a$, which blurs local irregularities and makes the output features smoother. After that, we use the time correlation module (TC, the TC module will be introduced in detail later) for dynamic learning. The final trend can be expressed as $\mathcal{X}^{t}:\{x^{t}_{u,t}\}_{u\in\mathcal{S},t\in[1,T]}=\alpha* Avg({x}_{u,t})+\beta*TC({x}_{u,t})$, where $\alpha$, $\beta$ denote learnable parameters. The fluctuation $\mathcal{X}^{f}:x^{f}_{u,t}$ can be obtained by subtracting $x^{t}_{u,t}$ from ${x}_{u,t}$.

\subsection{Correlation modeling}
\label{1.1}
\label{correlation}
\subsubsection{Fluctuation branch.}
In the fluctuation branch, in order to fully exploit fluctuation information for earnings forecasting, we perform correlation modeling on the time dimension of the same stock $u$ and on the stock dimension at the same time $t$.

%：以往的工作进行时间相关性建模主要有两种，一种是将多个时间维度的信息聚合为一个嵌入，由于不同股票间存在如产业上下游等导致的滞后影响问题，这种方法在进行股票间相关性建模会损失信息~\cite{master}将这种问题定义为跨时间相关性，因此master在建模时间相关性时保留了每个时间步的信息，但是【master】没有考虑到股票信息存在一种时间上的前后顺序关系及其因果性，即股票信息序列存在一个时间上的先后顺序，同时当前时间步的股票信息受过往时间步的信息影响，而与后续时间步信息无关。为了在建模时间相关性时能够满足这种时间先后性和因果性，具体来说

\begin{figure}[htbp]
\centering
     \includegraphics[width=0.9\linewidth]{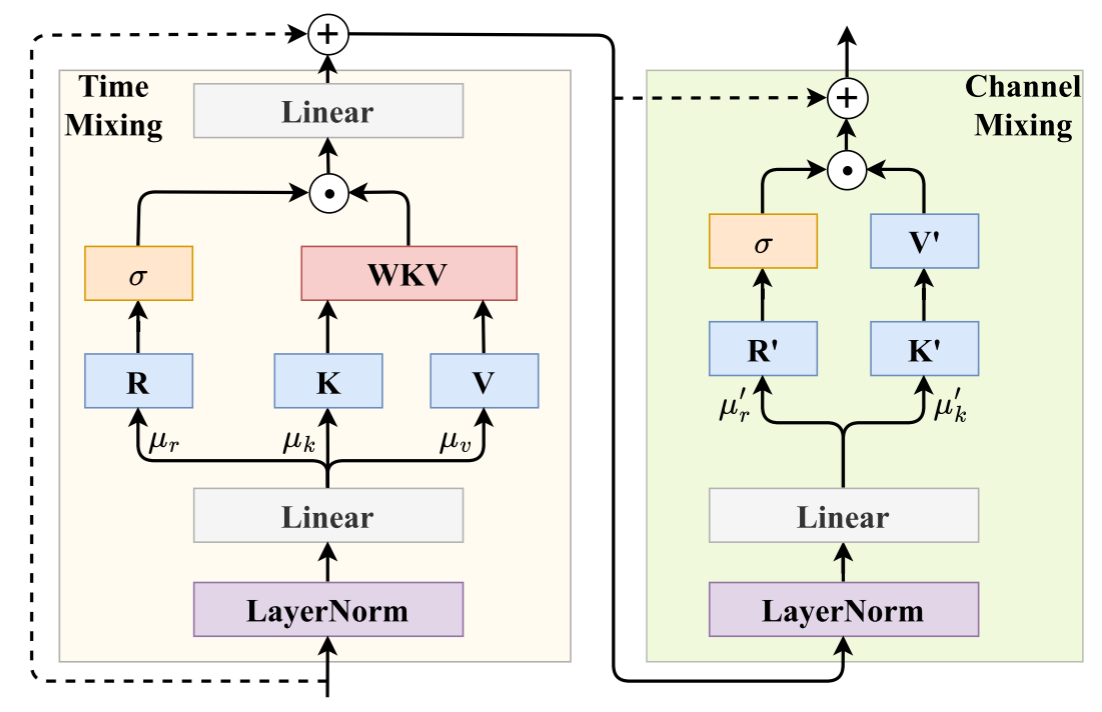}
\caption{The RWKV module structure.}
\label{fig:rwkv}
\end{figure}

\textbf{Time Correlation modeling (TC)} We not only retain the information of each time step, but also allow the first $t\in[1,T]$ time steps of each stock $u$ to interact with each other in order to meet the causal relationship of time modeling. We employ the RWKV model~\cite{peng2023rwkv} that combines the power of RNN and Transformer. The model absorbs the idea of RNN, so it can naturally maintain the time sequence of input features, and its internal linear attention mechanism also satisfies the causality of the input. The RWKV model structure is shown in Figure \ref{fig:rwkv}, which mainly includes Time-Mixing and Channel-Mixing blocks.
%，通过token移位保证了每个子块的输入是当前时刻和之前时刻的时间步特征。
The Time-Mixing block uses three learnable parameter matrices $\mu _r,\mu_k,\mu_v$ to balance the current input and the previous input. That is, $\mathbf{R}=\mu_rx,\mathbf{K}=\mu_kx,\mathbf{V}=\mu_vx$, where $\mathbf{R},\mathbf{K}$ and $\mathbf{V}$ represent the acceptance vector, key vector, and value vector, respectively. $\mathbf{WKV}$ in Figure \ref{fig:rwkv} is the linear attention operation in RWKV. The process can be expressed as: $$wkv_{t}=\frac{\sum_{i=1}^{t-1}e^{-(t-1-i) w+k_{i}}\odot v_{i}+e^{u+k_{t}}\odot v_{t}}{\sum_{i=1}^{t-1}e^{-(t-1-i) w+k_{i}}+e^{u+k_{t}}}.$$ where $w,u$ are two trainable parameters, $t$ is the current timestamp index, $\odot$ represents element-wise multiplication. In the numerator and denominator, $\sum_{i=1}^{t-1}e^{-(t-1-i) w+k_{i}}\odot v_{i}$ represents the information accumulation of $[1,t-1]$. Since $i < t$, the time step $t$ can only obtain the stock information before $t$, thus ensuring the temporal causality of the input information. Channel-Mixing block obtains the input of Time-Mixing block for information fusion again, and mixes different channel features through nonlinear mapping. The calculation methods of $\mathbf{R}^{\prime},\mathbf{K}^{\prime}$ and $\mathbf{V}^{\prime}$ are similar.
Finally, the original time embedding of the stock $\mathcal{X}^{f}$ is well modeled by RWKV, $H^{f}:{h^{f}_u}=\mathrm{RWKV}\left(x^{f}_{u}\right)\in \mathbb{R}^{T\times D}$, and each time step $t$ retains local details and encodes the information of historical time steps.

\textbf{Stock Correlation modeling (SC)} After modeling the correlation of stocks in the time dimension, we consider modeling the correlation between stocks at the same time step to provide additional information for the stocks. There is no temporal order and causality between different stocks, so we use the self-attention mechanism to model the correlation between stocks more flexibly. The stock fluctuation sequence $h^{f}_{t}\in \mathbb{R}^{S\times D}$ at each time step $t$ is used as the input sequence of the self-attention mechanism, and then the attention matrix $M\in\mathbb{R}^{S \times S}$ is obtained, so as to observe the correlation effect between any two stocks at different time steps. The process can be expressed as:
$$Q_{t}=W_{Q}H^{f}_{t},\quad K_{t}=W_{K}H^{f}_{t},\quad V_{t}=W_{V}H^{f}_{t},$$
$$Z^{f}_{t}=||_{u\in\mathcal{S}}z^{f}_{u,t}=\mathrm{FFN}(\mathrm{MHA}(Q_{t},K_{t},V_{t})+H^{f}_{t}),$$
where $W_Q, W_K, W_V$ are trainable parameters, $H^{f}_t=||_{u\in S}h^{f}_u$ represents local embedding of all stocks, FFN is a MLP with ReLU activation and residual connection, and MHA is the multi-head attention mechanisms.

The whole process of the fluctuation branch is as follows: first, the embedding of each stock’s changes is obtained through correlation modeling in the time dimension, and then the information of the mutual influence between stocks is obtained through correlation modeling in the stock dimension:
$$\{z^{f}_{u,t}\}_{u\in\mathcal{S},t\in[1,T]}=SC(TC(x^{f}_{u,t})).$$

\subsubsection{Trend branch.}
The trend branch is similar to the fluctuation branch, both of which include modeling the time correlation and stock correlation. The trend branch first models the correlation of stocks at the same time step because stock trends are more stable than their fluctuations. Recognizing the correlation between the trends of different stocks can provide mutual information, leading to a more accurate trend representation at the current time step. In addition, integrating the trend information of all stocks captures the overall market trend. Similar to fluctuations, trends also exhibit temporal order and causality, which is why we use RWKV to model time correlations. After two stages of correlation modeling, the trend embedding not only leverages the information between different stocks but also integrates trend information across the time dimension. The
operations are defined as follows: $$\{z^{t}_{u,t}\}_{u\in\mathcal{S},t\in[1,T]}=TC(SC(x^{t}_{u,t})).$$

\subsection{Output prediction}
Besides the trend and fluctuation embeddings, we also introduce market guidance information. Since the dimension of the market feature is small, we will first process it to obtain the information embedding $Z^m$ of the same dimension. We use a convolution kernel with $kernel = stride = k_c, padding = 0$ to perform a convolution operation, which is then linearly mapped to obtain $Z^{m}$.
This process can be formulated as:
$$Z=FFN(Z^{t}+Z^{f}+Z^{m}),$$
where $Z^{t}, Z^{f}$ and $Z^{m}$ represent the trend branch output, the fluctuation branch output, and the market guidance information embedding, respectively.
In order to generate the final stock representations, we use a method similar to MASTER~\cite{master} to aggregate the embeddings of stocks in T time steps, and use the embedding $z_{u,T}$ of the last time step T as the query vector to aggregate the attention score results:
$$\lambda_{u,t}=\frac{\exp(z_{u,t}^TW_\lambda z_{u,T})}{\sum_{i=1}^T\exp(z_{u,i}^TW_\lambda z_{u,T})}.$$
We use a linear layer as the predictor and evaluate the results by Mean Squared Error ($\textsf{MSE}$) loss function.
%In each batch, the model is jointly optimised for all $u \in \mathcal{S}$ on a particular prediction timestamp. A training epoch consists of multiple batches that correspond to different prediction dates in the training set. 
$$\hat{r}_u=FFN(\sum_{t\in[1,T]}\lambda_{u,t}z_{u,t}), \quad Loss=\sum_{u\in \mathcal{S}}\textsf{MSE}\left(r_u, \hat{r}_u \right).$$

\section{Experiments}

\subsubsection{Datasets.}
We evaluate our framework on the Chinese and US stock market with CSI300, CSI800 and S\&P500 stock datasets. CSI300 and CSI800 contain the 300 and 800 stocks with the highest market capitalization in the Shanghai and Shenzhen Stock Exchanges. S\&P500~\cite{stockmixer} includes the information about industries in the S\&P500 index from the Yahoo Finance database. The time range is from 2008.01.01 to 2023.12.31. The statistics of the datasets are summarised in Table~\ref{tab:dataset config}. We follow the experimental settings of MASTER~\cite{master} and construct sample features from the collected data using the publicly available Alpha158 indicators ~\cite{yang2020qlib}.
The lookback window length $T$ and prediction interval $d$ are set as $8$ and $5$, respectively. 
For market representation, we constructed $63$ features for CSI300 and CSI800 datasets with CSI300, CSI500 and CSI800 market indices . We include GSPC, DJI and NDX market indices for US market. The refereable interval $d'$ is set to $5,10,20,30,60$. 

\begin{table}[htbp]
\centering
\begin{tabular}{lccc}
\hline
Market            & CSI300              & CSI800      & S\&P500        \\ \hline
Stocks            & 300                 & 800         & 500        \\
Training period   & \multicolumn{3}{c}{2008/01/01-2020/03/31} \\
Validation period & \multicolumn{3}{c}{2020/04/01-2020/06/30} \\
Test period       & \multicolumn{3}{c}{2020/07/01-2023/12/31} \\
\# Features       & \multicolumn{3}{c}{158 + 63}                   \\ \hline
\end{tabular}
\caption{Detailed statistics of the datasets.}
\label{tab:datasetconfig}
\label{tab:dataset config}
\end{table}

\subsubsection{Baselines.}
We compare the performance of \frameworkname~ with several stock price forecasting baselines from different categories. 
$\bullet$ LSTM~\cite{lstm2017stock}: a Long Short-Term Memory network based stock price forecasting method. 
$\bullet$ Informer~\cite{informer}, DLinear~\cite{dlinear}, iTransformer~\cite{itransformer}, TimeMixer~\cite{timemixer}: state-of-the-art (SOTA) temporal prediction models using Transformer or MLP structure. \\
$\bullet$ DTML~\cite{dtml2021accurate}: an effective stock correlation mining method, which adopts the attention mechanism to mine the dynamic correlation among stocks and incorporates the market information into modeling.
$\bullet$ StockMixer~\cite{stockmixer}: a simple but powerful MLP-based architecture that performs indicator mixing, followed by time mixing, and finally stock mixing to imitate information exchange and market influences explicitly. 
$\bullet$ MASTER~\cite{master}: a state-of-the-art stock price forecasting model, which uses the transformer to model the momentary and cross-time stock correlation and leverages market information for automatic feature selection.

\begin{table*}[ht]
\centering
\small
\begin{tabular}{c|l|cccc|cc}
\toprule
 Dataset & Model & IC $\uparrow$& ICIR $\uparrow$ & RankIC $\uparrow$ & RankICIR $\uparrow$ & AR $\uparrow$ & IR $\uparrow$\\
 \midrule
\multirow{9}*{CSI300}
& LSTM 
&  $0.041\pm0.003$ & $0.29\pm0.04$ & $0.043\pm0.005$ & $0.30\pm0.04$ & $0.09\pm0.05$ & $0.9\pm0.5$ 
\\
& Informer  
& $0.040\pm0.008$ & $0.31\pm0.06$ & $0.037\pm0.010$ & $0.31\pm0.06$ & $0.14\pm0.03$ & $1.5\pm0.4$ 
\\
& DLinear 
& $0.040\pm0.001$ & $0.25\pm0.01$ & $0.037\pm0.001$ &  $0.26\pm0.01$ & $0.04\pm0.01$ & $0.4\pm0.1$ 
\\
& TimeMixer 
& $0.039\pm0.001$ & $0.26\pm0.01$ & $0.037\pm0.002$ & $0.26\pm0.01$ & $0.06\pm0.01$ & $0.6\pm0.1$ 
\\
& iTransformer 
& $0.043\pm0.001$ & $0.28\pm0.01$ & $0.042\pm0.003$ & $0.28\pm0.01$ & $0.07\pm0.01$ & $0.6\pm0.1$ 
\\ 
& DTML 
& $0.049\pm0.004$ & $0.34\pm0.02$ & $0.049\pm0.004$ & $0.34\pm0.02$ & $0.16\pm0.01$ & $1.6\pm0.2$ 
\\
& StockMixer 
& $0.041\pm0.006$ & $0.29\pm0.06$ & $0.041\pm0.008$ & $0.29\pm0.07$ & $0.09\pm0.08$ & $0.9\pm0.8$ 
\\
& MASTER 
& \underline{$0.055\pm0.002$} & \underline{$0.41\pm0.02$} & \underline{$0.054\pm0.002$} & \underline{$0.41\pm0.02$} & \underline{$0.20\pm0.02$} & \underline{$1.9\pm0.2$} 
\\
& Ours
& \bm{$0.143^*\pm0.001$} & \bm{$1.65^*\pm0.06$} & \bm{$0.097^*\pm0.001$} & \bm{$0.94^*\pm0.04$} & \bm{$0.80^*\pm0.06$} & \bm{$8.7^*\pm0.9$} 
\\
\midrule
\multirow{9}*{CSI800} 
& LSTM 
&  $0.043\pm0.004$ & $0.36\pm0.05$ & $0.043\pm0.006$ & $0.35\pm0.04$ & $0.15\pm0.04$ & $1.2\pm0.3$ 
\\
& Informer
& $0.043\pm0.004$ & $0.39\pm0.04$ & $0.047\pm0.006$ & $0.39\pm0.04$ & $0.18\pm0.04$ & $1.6\pm0.3$ 
\\
& DLinear
& $0.033\pm0.001$ & $0.27\pm0.01$ & $0.033\pm0.001$ & $0.27\pm0.01$ & $0.06\pm0.01$ & $0.5\pm0.1$ 
\\
& TimeMixer
& $0.033\pm0.001$ & $0.29\pm0.01$ & $0.033\pm0.003$ & $0.28\pm0.01$ & $0.05\pm0.01$ & $0.4\pm0.1$ 
\\
& iTransformer
& $0.034\pm0.001$ & $0.28\pm0.01$ & $0.034\pm0.001$ & $0.28\pm0.01$ & $0.07\pm0.01$ & $0.6\pm0.1$ 
\\
& DTML
& $0.044\pm0.001$ & $0.35\pm0.02$ & $0.049\pm0.003$ & $0.35\pm0.02$ & $0.14\pm0.01$ & $1.1\pm0.1$ 
\\
& StockMixer
& $0.032\pm0.002$ & $0.27\pm0.03$ & $0.034\pm0.002$ & $0.27\pm0.02$ & $0.06\pm0.02$ & $0.5\pm0.1$ 
\\ 
& MASTER
& \underline{$0.048\pm0.001$} & \underline{$0.44\pm0.03$} & \underline{$0.051\pm0.003$} & \underline{$0.44\pm0.02$} & \underline{$0.19\pm0.02$} & \underline{$1.6\pm0.1$} 
\\ 
& Ours 
& \bm{$0.138^*\pm0.007$} & \bm{$1.54^*\pm0.62$} & \bm{$0.092^*\pm0.003$} & \bm{$0.88^*\pm0.08$} & \bm{$1.75^*\pm0.22$} & \bm{$14.2^*\pm1.8$} 
\\
 \midrule
\multirow{9}*{S\&P500}
& LSTM 
&  $0.006\pm0.001$ & $0.04\pm0.01$ & $0.005\pm0.001$ & $0.04\pm0.01$ & \underline{$0.09\pm0.03$} & \underline{$0.8\pm0.2$} 
\\
& Informer  
& $0.015\pm0.001$ & $0.08\pm0.01$ & $0.013\pm0.002$ & $0.09\pm0.01$ & $0.07\pm0.02$ & $0.6\pm0.2$
\\
& DLinear 
& $0.013\pm0.001$ & $0.09\pm0.01$ & $0.012\pm0.001$ &  $0.09\pm0.01$ & $0.08\pm0.01$ & $0.7\pm0.1$ 
\\
& TimeMixer
& $0.014\pm0.001$ & $0.10\pm0.02$ & $0.013\pm0.001$ & $0.10\pm0.02$ & $0.07\pm0.01$ & $0.7\pm0.1$ 
\\ 
& iTransformer 
& $0.014\pm0.001$ & $0.09\pm0.01$ & $0.013\pm0.001$ & $0.09\pm0.01$ & $0.09\pm0.02$ & $0.7\pm0.2$ 
\\
& DTML 
& $0.012\pm0.001$ & $0.10\pm0.01$ & $0.011\pm0.001$ & $0.11\pm0.01$ & $0.05\pm0.02$ & $0.5\pm0.1$ 
\\
& StockMixer 
& $0.013\pm0.001$ & $0.10\pm0.01$ & $0.012\pm0.001$ & $0.09\pm0.04$ & $0.06\pm0.01$ & $0.6\pm0.1$ 
\\
& MASTER 
& \underline{$0.021\pm0.001$} & \underline{$0.14\pm0.01$} & \underline{$0.019\pm0.001$} & \underline{$0.15\pm0.01$} & $0.08\pm0.02$ & $0.7\pm0.1$ 
\\
& Ours
& \bm{$0.114^*\pm0.009$} & \bm{$1.11^*\pm0.19$} & \bm{$0.053^*\pm0.006$} & \bm{$0.60^*\pm0.15$} & \bm{$0.84^*\pm0.10$} & \bm{$8.2^*\pm1.2$}
\\
\bottomrule
\end{tabular}
\caption{Overall performance comparison. The best results are in bold and the second-best results are underlined. And * denotes
statistically significant improvement (measured by t-test with p-value $<$ 0.01) over all baselines.}
\label{tab:Different models performance}
\end{table*}

\subsubsection{Implementation.}
We implement the model using PyTorch and follow the data settings of previous work~\cite{master}. The model is trained using CosineAnnealingLR ~\cite{loshchilov2016sgdr} with a warm-up phase and optimized by the Adam optimizer. The learning rate is kept in the range $\left(2 \times 10^{-4}, 3 \times 10^{-3}\right)$. The warm-up epochs are set to $10$ and the number of epochs between two restarts is $15$. A total of $75$ epochs are trained and the results of the last epoch are selected for testing. All experiments are conduct on a server equipped with an Intel(R) Xeon(R) CPU E5-2678 v3 (48 CPUs, 125GB) and a NVIDIA GeForce RTX 3090 GPU (24 GB). We run the training and testing procedures 5 times, and the average and standard deviation are reported for all methods.

\subsubsection{Metrics.}
To comprehensively measure the performance of the framework, we employ ranking metrics and return on investment metrics based on previous work~\cite{master}. IC and RankIC (RIC) are the Pearson coefficient and Spearman coefficient averaged at a daily frequency. ICIR and RankICIR (RICIR) are normalized metrics of IC and RankIC by dividing the standard deviation. The above four metrics evaluate stock selection ability and overall ranking performance of model. We also use investment-based indicators AR and IR. AR measures the annual expected excess return generated by the investment, while IR measures the risk-adjusted performance of an investment.

% \begin{figure}[htbp]
% \centering
% \small
%      \includegraphics[width=1.0\linewidth]{backtest.png}
% \caption{Cumulative return. Benchmark represents the CSI300 and S\&P500 market indices, respectively.}
% \label{fig:stock2date_sp500}
% \end{figure}

\subsection{Overall Comparison}
Table~\ref{tab:Different models performance} reports the overall performance. On three representative stock datasets, \frameworkname~has achieved an absolute lead compared with both the classic time series prediction model and the latest stock price prediction SOTA. \frameworkname~has a $250\% \sim 500\%$ improvement over SOTA~\cite{master} in ranking-based indicators and a $400\% \sim 1000\%$ improvement in portfolio-based indicators. By visualizing the metrics in Figure~\ref{fig:spider1}, it is clear that our model outperforms other baselines by a wide margin. It is worth noting that while MASTER, DTML and StockMixer also design their own correlation modeling modules, our model still achieves the best performance. We also find that although DLinear and TimeMixer employ the decomposition block in their framework, they still fail to perform well. We attribute this improvement to the trend and fluctuation dual-branch framework, which independently models the strong correlation between two different types of information after effectively decomposing them. 

\begin{figure}[htbp]
\centering
\small
     \includegraphics[width=1.0\linewidth]{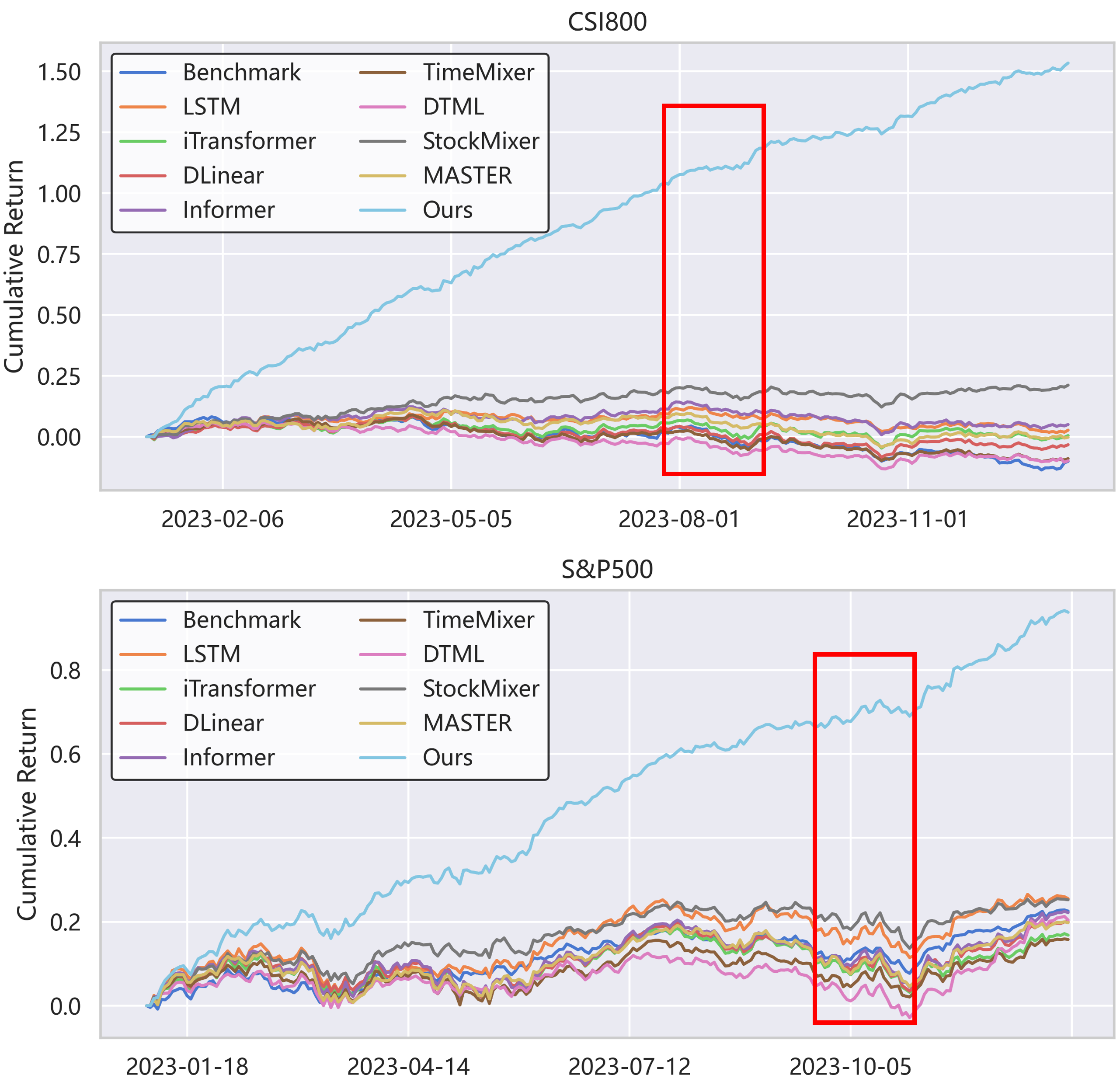}
\caption{Cumulative portfolio returns on the CSI800 and S\&P500 test sets. Benchmark represents the CSI300 and S\&P500 market indices, respectively.}
\label{fig:stock2date_sp500}
\end{figure}

In order to verify the profitability of the model, we conduct a backtest experiment during the test dataset period as well. Following previous works~\cite{master, stockformer}, we adopt "Top30-Drop30" strategy to simulate daily trading. "Top30" means keeping the stocks with the top 30 return ratios; and "Drop30" means dropping any stock whose score falls out of the top 30, regardless of its previous performance. As shown in Figure~\ref{fig:stock2date_sp500},  within the red box area, when the benchmark indicates an overall decline, most models also exhibit negative returns. However, our model consistently maintains a high positive return during the whole test periods.

\begin{table}[h]
\small
\centering
\begin{tabular}{l|cccc}
\toprule                 & IC    & ICIR  & RIC & RICIR \\ 
\midrule
 w/o Decomposition       & $0.045$ & $0.43$ & $0.047$  & 0.44    \\
 w/o Fluctuation & $0.050$ & $0.44$ & $0.052$  & $0.45$    \\
 w/o Trend    & $0.134$ & $1.53$ & $0.086$  & $0.87$    \\
 w/o SC       & $0.071$ & $0.75$ & $0.057$  & $0.46$    \\
 w/o TC              & $0.097$ & $0.93$ & $0.077$  & $0.69$    \\
 \midrule
 TC-SA         & $0.121$ & $1.36$ & $0.088$  & $0.81$    \\
 TC-maskSA       & $0.131$ & $1.40$ & $0.091$  & $0.84$    \\
 Ours  & \bm{$0.138$} & \bm{$1.54$}& \bm{$0.092$}  &  \bm{$0.88$} \\ 
 MASTER-modified & $0.128$ & $1.38$& $0.088$ &  $0.84$ \\ 
\bottomrule
\end{tabular}
\caption{Ablation experiments on CSI800 to validate the effectiveness of the proposed dual-branch decomposition framework and modeling temporal causality mechanism. The best results are in bold.}
\label{table:ablation}
% \vspace{-10px}
\end{table}

\subsection{Ablation Study}
To validate the design choices in our proposed framework, we perform an ablation experiment by removing components individually: Decomposition (w/o Decomposition), correlations in Fluctuation branch (w/o Fluctuation), correlations in Trend branch (w/o Trend), Stock Correlation (w/o SC) and Time Correlation (w/o TC). In order to verify the effectiveness of our temporal causality mechanism, we use self attention (TC-SA) and masked self attention (TC-maskSA) to model time correlation instead of RWKV.
The experiments are performed on the CSI800 dataset, and the results are presented in Table~\ref{table:ablation}. We observe the following points: 
\begin{itemize}
\item When the Decomposition module is removed, the performance of the variant drops significantly, which demonstrates the necessity of using dual branches to decompose information because it can avoid interference between the two types of information. 
\item The absence of correlation modeling in either the trend branch or the fluctuation branch results in decreased model performance, with a more significant drop observed in the latter. This indicates that fluctuation information requires more extensive correlation analysis than trend information.
\item The model performance drops when either the time correlation module (TC) or the stock correlation module (SC) is removed, indicating that both time dependence and stock dependence are significant factors in the stock market. 
\item Owing to masking future information, improved results are achieved in both TC-maskSA model and ours, which demonstrates the effectiveness of the temporal causality discussed in this paper for predicting stock prices. 
\item Notably, we modify the MASTER structure by incorporating a plug-and-play decomposition block and adopting a dual-branch structure. As shown in MASTER-modified, the performance is improved by more than two times compared with the original MASTER, which reinforces the generalized ability and effectiveness of the proposed blocks.
\end{itemize}

We also conducted experiments to verify the effect of the order of correlation modeling on the trend and fluctuation branches. $\mathcal{TCSC}$ stands for modeling time correlation first, then stock correlation, and $\mathcal{SCTC}$ the other way around. The results are shown in Table~\ref{tab:correlation orders}.

After adjusting the order, the metrics of both model variants decrease, including a serious decline after replacing the order of the fluctuation branch, which proves that if the volatility information is first modeled on the stock dimension, it will lead to confusion about the volatility information of different stocks and increase the difficulty of model learning. Trend information, on the other hand, is more stable and better suited to prioritize stock information interactions for comprehensive information. This experiment also confirms our original intention that the trend is a centralized reflection of the whole stock movement and fluctuation is a unique feature of individual stock changes.

% \begin{figure}[htbp]
% \centering
% \small
%      \includegraphics[width=1.0\linewidth]{bacttest.png}
% \caption{Cumulative return. Benchmark represents the CSI300 and S\&P500 indices, respectively.}
% \label{fig:stock2date_sp500}
% \end{figure}

% \begin{figure}[htbp]
% \centering
%      \includegraphics[width=1.0\linewidth]{Hyperparameter_sensitivity.png}
% \caption{Hyper-parameters sensitivity}
% \label{fig:seq_len}
% \end{figure}

% \begin{figure}[htbp]
% \centering
%      \includegraphics[width=1.0\linewidth]{NEW_WORK/New_Work_CSI800_202301_202312.png}
% \caption{Cumulative return. Benchmark 代表CSI300指数}
% \label{fig:stock2date}
% \end{figure}

% \begin{figure}[htbp]
% \centering
%      \includegraphics[width=1.0\linewidth]{NEW_WORK/New_Work_SP500_202301_202312.png}
% \caption{Cumulative return. Benchmark 代表S\&P500指数}
% \label{fig:stock2date_sp500}
% \end{figure}
\begin{table}[htbp]
    \centering
    \small
    \setlength{\tabcolsep}{1.1mm}
\begin{tabular}{cccc|cccc}
\toprule
\multicolumn{2}{c}{Fluctuation} & \multicolumn{2}{c|}{Trend} & \multirow{2}{*}{IC} & \multirow{2}{*}{ICIR} & \multirow{2}{*}{RIC} & \multirow{2}{*}{RICIR} \\
$\mathcal{TCSC}$           & $\mathcal{SCTC}$           & $\mathcal{TCSC}$         & $\mathcal{SCTC}$        &                     &                       &                      &                        \\ \midrule
               -&                \checkmark&              -&             \checkmark&                     0.120&                       1.28&                      0.082&                        0.81\\
               -&                \checkmark&              \checkmark&             -&                     0.114&                       1.24&                      0.079&                        0.79\\
               \checkmark&                -&              \checkmark&             -&                     0.136&                       1.54&                      0.089&                        0.84\\
               \checkmark&                -&              -&             \checkmark&             \bm{$0.138$} &                \bm{$1.54$}&               \bm{$0.092$}&                 \bm{$0.88$}\\ \bottomrule
\end{tabular}
\caption{\frameworkname~performance with varying correlation orders. The best results are in bold.}
\label{tab:correlation orders}
\vspace{-10px}
\end{table}

\subsection{Hyperparameter Sensitivity}
We have performed hyperparameter sensitivity experiments on the CSI800 dataset to explore the effect of different settings on our framework performance in Figure~\ref{fig:seq_len}. We focus on the Avgpool kernel $k_a$, the market convolution kernel $k_c$, the RWKV heads, the lookback window $T$, and the prediction interval $d$. IC and RIC are selected as the metrics.
\subsubsection{Avgpool kernel.}
% We first study the effect of different Avgpool kernel sizes $k_a$ on the decomposition of the trend and fluctuation components. 
As shown in Figure~\ref{fig:seq_len}(a), we find that larger pool kernels $k_a$ perform better because the model is able to aggregate trend information over a longer period of time. Therefore, the extracted trend information is more representative of the long-term movement of the stocks.
\subsubsection{Market Convolution kernel.}
% We consider the impact of different convolution kernel sizes $k_c$ on market information. 
We observe that smaller convolution kernels $k_c$ can only focus on a short period of time, and the market often changes dramatically; when the convolution kernel is large, it often leads to blurring and loss of previous market information.
\subsubsection{RWKV head.}
We find that an appropriate number of heads can fully utilize representation embedding, while too many or too small heads reduce information utilization.
\subsubsection{Lookback window.}
\begin{figure}[htbp]
\centering
     \includegraphics[width=1.0\linewidth]{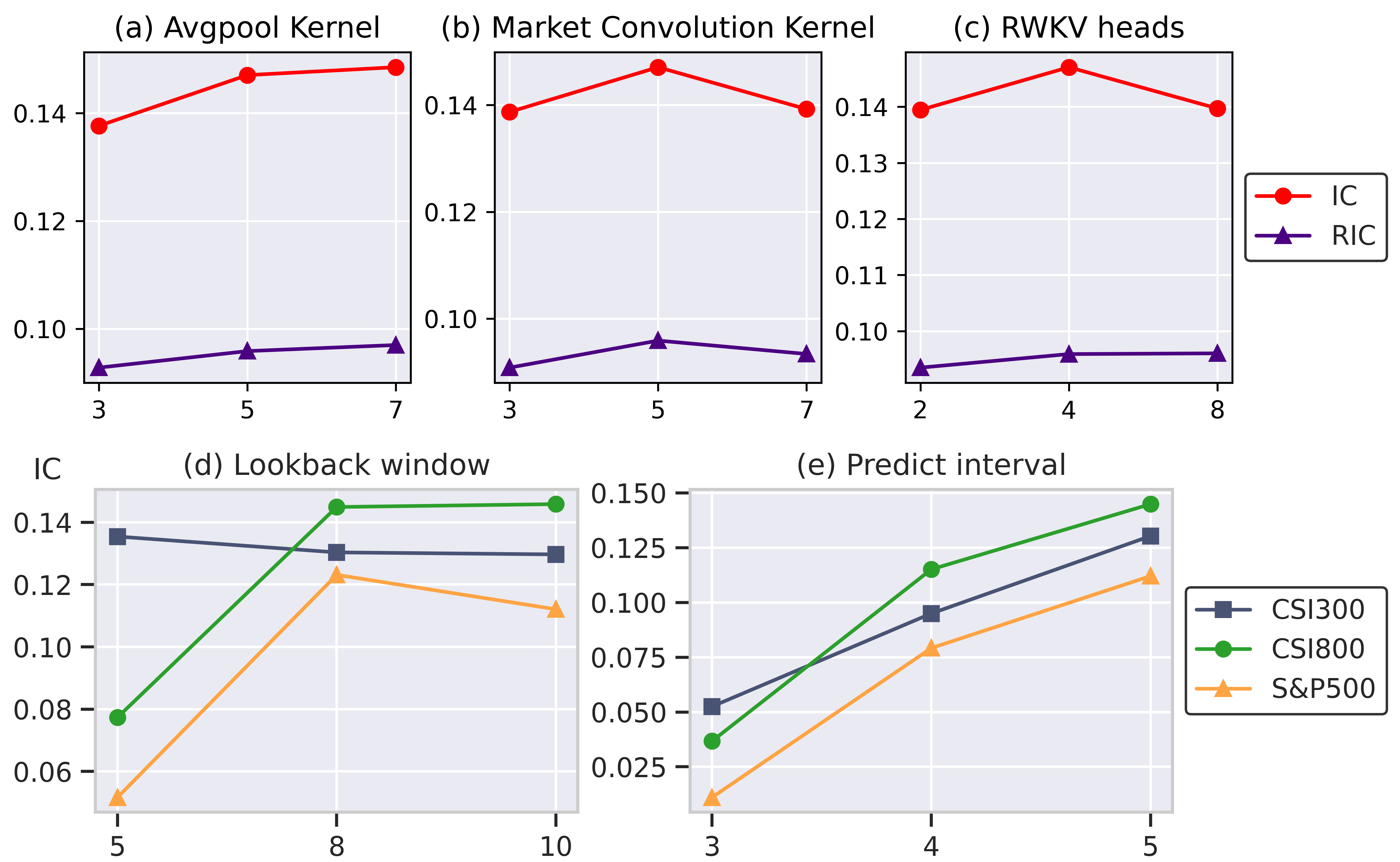}
\caption{The results of hyperparameter sensitivity. The x-axis in the plots represents Avgpool kernel size, convolutional kernel size, RWKV heads, lookback window length, and predict interval, respectively. The results of (a), (b) and (c) are obtained in the CSI800 dataset.}
\label{fig:seq_len}
\vspace{-10px}
\end{figure}

We also explore the effect of different lookback window lengths $T$ on the predictive effectiveness of the model. As shown in Figure~\ref{fig:seq_len}(d), for most of the datasets, the model tends to perform better when the lookback window is longer because it has access to more historical information to capture trends and changes in volatility.
\subsubsection{Prediction interval.}
% We analyze the impact of the prediction length $d$ on the model performance. 
A shorter prediction length makes it difficult to learn random labels, resulting in poor prediction results. Properly increasing the prediction length can make the model perform better.

\section{Discussion}
Due to diverse stock trading mechanisms and market rules in different countries, how to fully capture the market information requires more granular modeling of the unique trading features of each market. Considering the limited space and resources, we focus our experiments on Chinese and US stock market. We also find that the indicators are greatly affected by extreme samples. If the stock data with extreme returns are retained during data preprocessing, the evaluation results of most models are greatly affected. Therefore, we use a method similar to previous work~\cite{gao2024diffsformer} to handle the data.

\section{Conclusion}
We introduce a new framework for stock price prediction, \frameworkname, which re-examines the composition and importance of stock information, decomposes it into trend and fluctuation components, and uses independent correlation modeling branches to process it. At the same time, considering that the fluctuation component is a separate attribute of a stock, while the trend can be jointly reflected by the stock market, the opposite correlation modeling process is set up, and the causal relationship of time series data such as stocks is innovatively considered. On the three mainstream data sets, compared with all benchmarks, \frameworkname~has a huge improvement in all four indicators compared to the previous SOTA model. In short, \frameworkname~proposes a new perspective for stock price prediction. In the future, our goal is to study stock price prediction with finer granularity in time and explore greater possibilities for high-frequency trading.

\section{Acknowledgments}
% (At the same time, in order to prove the difference in stock relationships at different time steps, we visualized the attention matrix)
The authors would like to thank the anonymous reviewers for their insightful reviews.

\nobibliography*
\bibliography{aaai25}

\begin{thebibliography}{29}
\providecommand{\natexlab}[1]{#1}

\bibitem[{Bai, Kolter, and Koltun(2018)}]{conv2018empirical}
Bai, S.; Kolter, J.~Z.; and Koltun, V. 2018.
\newblock An empirical evaluation of generic convolutional and recurrent networks for sequence modeling.
\newblock \emph{arXiv preprint arXiv:1803.01271}.

\bibitem[{Chen and Guestrin(2016)}]{chen2016xgboost}
Chen, T.; and Guestrin, C. 2016.
\newblock Xgboost: A scalable tree boosting system.
\newblock In \emph{Proceedings of the 22nd acm sigkdd international conference on knowledge discovery and data mining}, 785--794.

\bibitem[{Cho et~al.(2014)Cho, Van~Merri{\"e}nboer, Gulcehre, Bahdanau, Bougares, Schwenk, and Bengio}]{gru2014learning}
Cho, K.; Van~Merri{\"e}nboer, B.; Gulcehre, C.; Bahdanau, D.; Bougares, F.; Schwenk, H.; and Bengio, Y. 2014.
\newblock Learning phrase representations using RNN encoder-decoder for statistical machine translation.
\newblock \emph{arXiv preprint arXiv:1406.1078}.

\bibitem[{Ding et~al.(2020)Ding, Wu, Sun, Guo, and Guo}]{ding2020hierarchical}
Ding, Q.; Wu, S.; Sun, H.; Guo, J.; and Guo, J. 2020.
\newblock Hierarchical Multi-Scale Gaussian Transformer for Stock Movement Prediction.
\newblock In \emph{IJCAI}, 4640--4646.

\bibitem[{Fan and Shen(2024)}]{stockmixer}
Fan, J.; and Shen, Y. 2024.
\newblock StockMixer: A Simple Yet Strong MLP-Based Architecture for Stock Price Forecasting.
\newblock In \emph{Proceedings of the AAAI Conference on Artificial Intelligence}, volume~38, 8389--8397.

\bibitem[{Feng et~al.(2019)Feng, He, Wang, Luo, Liu, and Chua}]{feng2019temporal}
Feng, F.; He, X.; Wang, X.; Luo, C.; Liu, Y.; and Chua, T.-S. 2019.
\newblock Temporal relational ranking for stock prediction.
\newblock \emph{ACM Transactions on Information Systems (TOIS)}, 37(2): 1--30.

\bibitem[{Gao, Wang, and Yang(2023)}]{stockformer}
Gao, S.; Wang, Y.; and Yang, X. 2023.
\newblock StockFormer: Learning Hybrid Trading Machines with Predictive Coding.
\newblock In \emph{IJCAI}, 4766--4774.

\bibitem[{Gao et~al.(2024)Gao, Chen, Wang, Wang, Wang, Gao, and Ding}]{gao2024diffsformer}
Gao, Y.; Chen, H.; Wang, X.; Wang, Z.; Wang, X.; Gao, J.; and Ding, B. 2024.
\newblock DiffsFormer: A Diffusion Transformer on Stock Factor Augmentation.
\newblock \emph{arXiv preprint arXiv:2402.06656}.

\bibitem[{Huynh et~al.(2023)Huynh, Nguyen, Nguyen, Nguyen, Weidlich, Nguyen, and Aberer}]{huynh2023efficient}
Huynh, T.~T.; Nguyen, M.~H.; Nguyen, T.~T.; Nguyen, P.~L.; Weidlich, M.; Nguyen, Q. V.~H.; and Aberer, K. 2023.
\newblock Efficient integration of multi-order dynamics and internal dynamics in stock movement prediction.
\newblock In \emph{Proceedings of the Sixteenth ACM International Conference on Web Search and Data Mining}, 850--858.

\bibitem[{Kamble(2017)}]{kamble2017short}
Kamble, R.~A. 2017.
\newblock Short and long term stock trend prediction using decision tree.
\newblock In \emph{2017 International Conference on Intelligent Computing and Control Systems (ICICCS)}, 1371--1375. IEEE.

\bibitem[{Li et~al.(2024)Li, Liu, Shen, Wang, Chen, and Huang}]{master}
Li, T.; Liu, Z.; Shen, Y.; Wang, X.; Chen, H.; and Huang, S. 2024.
\newblock MASTER: Market-Guided Stock Transformer for Stock Price Forecasting.
\newblock In \emph{Proceedings of the AAAI Conference on Artificial Intelligence}, volume~38, 162--170.

\bibitem[{Li et~al.(2020)Li, Bao, Harimoto, Chen, Xu, and Su}]{ijcai2020p626}
Li, W.; Bao, R.; Harimoto, K.; Chen, D.; Xu, J.; and Su, Q. 2020.
\newblock Modeling the Stock Relation with Graph Network for Overnight Stock Movement Prediction.
\newblock In Bessiere, C., ed., \emph{Proceedings of the Twenty-Ninth International Joint Conference on Artificial Intelligence, {IJCAI-20}}, 4541--4547. International Joint Conferences on Artificial Intelligence Organization.
\newblock Special Track on AI in FinTech.

\bibitem[{Liu et~al.(2019)Liu, Lin, Liu, Xu, Ren, Diao, and Yang}]{liu2019transformer}
Liu, J.; Lin, H.; Liu, X.; Xu, B.; Ren, Y.; Diao, Y.; and Yang, L. 2019.
\newblock Transformer-based capsule network for stock movement prediction.
\newblock In \emph{Proceedings of the first workshop on financial technology and natural language processing}, 66--73.

\bibitem[{Liu et~al.(2023)Liu, Hu, Zhang, Wu, Wang, Ma, and Long}]{itransformer}
Liu, Y.; Hu, T.; Zhang, H.; Wu, H.; Wang, S.; Ma, L.; and Long, M. 2023.
\newblock iTransformer: Inverted Transformers Are Effective for Time Series Forecasting.
\newblock \emph{arXiv preprint arXiv:2310.06625}.

\bibitem[{Loshchilov and Hutter(2016)}]{loshchilov2016sgdr}
Loshchilov, I.; and Hutter, F. 2016.
\newblock Sgdr: Stochastic gradient descent with warm restarts.
\newblock \emph{arXiv preprint arXiv:1608.03983}.

\bibitem[{Nelson, Pereira, and De~Oliveira(2017)}]{lstm2017stock}
Nelson, D.~M.; Pereira, A.~C.; and De~Oliveira, R.~A. 2017.
\newblock Stock market's price movement prediction with LSTM neural networks.
\newblock In \emph{2017 International joint conference on neural networks (IJCNN)}, 1419--1426. Ieee.

\bibitem[{Nugroho, Adji, and Fauziati(2014)}]{nugroho2014decision}
Nugroho, F. S.~D.; Adji, T.~B.; and Fauziati, S. 2014.
\newblock Decision support system for stock trading using multiple indicators decision tree.
\newblock In \emph{2014 The 1st International Conference on Information Technology, Computer, and Electrical Engineering}, 291--296. IEEE.

\bibitem[{Peng et~al.(2023)Peng, Alcaide, Anthony, Albalak, Arcadinho, Cao, Cheng, Chung, Grella, GV et~al.}]{peng2023rwkv}
Peng, B.; Alcaide, E.; Anthony, Q.; Albalak, A.; Arcadinho, S.; Cao, H.; Cheng, X.; Chung, M.; Grella, M.; GV, K.~K.; et~al. 2023.
\newblock Rwkv: Reinventing rnns for the transformer era.
\newblock \emph{arXiv preprint arXiv:2305.13048}.

\bibitem[{Qian et~al.(2024)Qian, Zhou, Zhao, Chen, Yao, Wang, Liu, Yu, Zhang, and Zhou}]{mdgnn}
Qian, H.; Zhou, H.; Zhao, Q.; Chen, H.; Yao, H.; Wang, J.; Liu, Z.; Yu, F.; Zhang, Z.; and Zhou, J. 2024.
\newblock MDGNN: Multi-Relational Dynamic Graph Neural Network for Comprehensive and Dynamic Stock Investment Prediction.
\newblock In \emph{Proceedings of the AAAI Conference on Artificial Intelligence}, volume~38, 14642--14650.

\bibitem[{Sawhney et~al.(2020)Sawhney, Agarwal, Wadhwa, and Shah}]{sawhney2020spatiotemporal}
Sawhney, R.; Agarwal, S.; Wadhwa, A.; and Shah, R.~R. 2020.
\newblock Spatiotemporal hypergraph convolution network for stock movement forecasting.
\newblock In \emph{2020 IEEE International Conference on Data Mining (ICDM)}, 482--491. IEEE.

\bibitem[{Wang et~al.(2024)Wang, Wu, Shi, Hu, Luo, Ma, Zhang, and ZHOU}]{timemixer}
Wang, S.; Wu, H.; Shi, X.; Hu, T.; Luo, H.; Ma, L.; Zhang, J.~Y.; and ZHOU, J. 2024.
\newblock TimeMixer: Decomposable Multiscale Mixing for Time Series Forecasting.
\newblock In \emph{International Conference on Learning Representations (ICLR)}.

\bibitem[{Xia et~al.(2024)Xia, Ao, Li, Liu, Liu, Ye, and Chai}]{ci-sthpan}
Xia, H.; Ao, H.; Li, L.; Liu, Y.; Liu, S.; Ye, G.; and Chai, H. 2024.
\newblock CI-STHPAN: Pre-trained Attention Network for Stock Selection with Channel-Independent Spatio-Temporal Hypergraph.
\newblock In \emph{Proceedings of the AAAI Conference on Artificial Intelligence}, volume~38, 9187--9195.

\bibitem[{Xiang et~al.(2022)Xiang, Cheng, Shang, Zhang, and Liang}]{xiang2022temporal}
Xiang, S.; Cheng, D.; Shang, C.; Zhang, Y.; and Liang, Y. 2022.
\newblock Temporal and Heterogeneous Graph Neural Network for Financial Time Series Prediction.
\newblock In \emph{Proceedings of the 31st ACM International Conference on Information \& Knowledge Management}, 3584--3593.

\bibitem[{Xie et~al.(2013)Xie, Passonneau, Wu, and Creamer}]{xie2013semantic}
Xie, B.; Passonneau, R.; Wu, L.; and Creamer, G.~G. 2013.
\newblock Semantic frames to predict stock price movement.
\newblock In \emph{Proceedings of the 51st annual meeting of the association for computational linguistics}, 873--883.

\bibitem[{Yang et~al.(2020)Yang, Liu, Zhou, Bian, and Liu}]{yang2020qlib}
Yang, X.; Liu, W.; Zhou, D.; Bian, J.; and Liu, T.-Y. 2020.
\newblock Qlib: An ai-oriented quantitative investment platform.
\newblock \emph{arXiv preprint arXiv:2009.11189}.

\bibitem[{Yoo et~al.(2021{\natexlab{a}})Yoo, Soun, Park, and Kang}]{dtml2021accurate}
Yoo, J.; Soun, Y.; Park, Y.-c.; and Kang, U. 2021{\natexlab{a}}.
\newblock Accurate multivariate stock movement prediction via data-axis transformer with multi-level contexts.
\newblock In \emph{Proceedings of the 27th ACM SIGKDD Conference on Knowledge Discovery \& Data Mining}, 2037--2045.

\bibitem[{Yoo et~al.(2021{\natexlab{b}})Yoo, Soun, Park, and Kang}]{yoo2021accurate}
Yoo, J.; Soun, Y.; Park, Y.-c.; and Kang, U. 2021{\natexlab{b}}.
\newblock Accurate multivariate stock movement prediction via data-axis transformer with multi-level contexts.
\newblock In \emph{Proceedings of the 27th ACM SIGKDD Conference on Knowledge Discovery \& Data Mining}, 2037--2045.

\bibitem[{Zeng et~al.(2023)Zeng, Chen, Zhang, and Xu}]{dlinear}
Zeng, A.; Chen, M.; Zhang, L.; and Xu, Q. 2023.
\newblock Are transformers effective for time series forecasting?
\newblock In \emph{Proceedings of the AAAI conference on artificial intelligence}, volume~37, 11121--11128.

\bibitem[{Zhou et~al.(2021)Zhou, Zhang, Peng, Zhang, Li, Xiong, and Zhang}]{informer}
Zhou, H.; Zhang, S.; Peng, J.; Zhang, S.; Li, J.; Xiong, H.; and Zhang, W. 2021.
\newblock Informer: Beyond Efficient Transformer for Long Sequence Time-Series Forecasting.
\newblock In \emph{The Thirty-Fifth {AAAI} Conference on Artificial Intelligence, {AAAI} 2021, Virtual Conference}, volume~35, 11106--11115. {AAAI} Press.

\end{thebibliography}

\section{Reproducibility Checklist}

This paper:
\begin{itemize}
\item Includes a conceptual outline and/or pseudocode description of AI methods introduced. (yes)
\item Clearly delineates statements that are opinions, hypothesis, and speculation from objective facts and results. (yes)
\item Provides well marked pedagogical references for less-familiare readers to gain background necessary to replicate the paper. (yes)
\end{itemize}

\noindent Does this paper make theoretical contributions? (no)

\noindent Does this paper rely on one or more datasets? (yes)

\begin{itemize}
\item A motivation is given for why the experiments are conducted on the selected datasets. (yes)
\item All novel datasets introduced in this paper are included in a data appendix. (partial)
\item All novel datasets introduced in this paper will be made publicly available upon publication of the paper with a license that allows free usage for research purposes. (yes)
\item All datasets drawn from the existing literature (potentially including authors’ own previously published work) are accompanied by appropriate citations. (yes)
\item All datasets drawn from the existing literature (potentially including authors’ own previously published work) are publicly available. (yes)
\item All datasets that are not publicly available are described in detail, with explanation why publicly available alternatives are not scientifically satisficing. (yes)
\end{itemize}
\noindent Does this paper include computational experiments? (yes)
\begin{itemize}
\item Any code required for pre-processing data is included in the appendix. (yes).
\item All source code required for conducting and analyzing the experiments is included in a code appendix. (yes)
\item All source code required for conducting and analyzing the experiments will be made publicly available upon publication of the paper with a license that allows free usage for research purposes. (yes)
\item All source code implementing new methods have comments detailing the implementation, with references to the paper where each step comes from. (yes)
\item If an algorithm depends on randomness, then the method used for setting seeds is described in a way sufficient to allow replication of results. (yes)
\item This paper specifies the computing infrastructure used for running experiments (hardware and software), including GPU/CPU models; amount of memory; operating system; names and versions of relevant software libraries and frameworks. (yes)
\item This paper formally describes evaluation metrics used and explains the motivation for choosing these metrics. (yes)
\item This paper states the number of algorithm runs used to compute each reported result. (yes)
\item Analysis of experiments goes beyond single-dimensional summaries of performance (e.g., average; median) to include measures of variation, confidence, or other distributional information. (yes)
\item The significance of any improvement or decrease in performance is judged using appropriate statistical tests (e.g., Wilcoxon signed-rank). (yes)
\item This paper lists all final (hyper-)parameters used for each model/algorithm in the paper’s experiments. (yes)
\item This paper states the number and range of values tried per (hyper-) parameter during development of the paper, along with the criterion used for selecting the final parameter setting. (yes)
\end{itemize}

\end{document}